\documentstyle[aps,prl,epsf,preprint]{revtex}
\input epsf.tex
\begin{document}
\title{QCD Phase Shifts and Rising Total Cross-Sections}
\author{C. S. Lam\cite{CSL}}
\address{Department of Physics, McGill University\\
3600 University St., Montreal, P.Q., Canada H3A 2T8}
\maketitle

\def\e{\epsilon}
\def\.{\cdot}
\def\o{\omega}
\def\be{\begin{eqnarray}}
\def\nn{\nonumber\\}
\def\ee{\end{eqnarray}}
\def\({\left(}
\def\[{\left[}
\def\){\right)}
\def\]{\right]}
\def\t{\tau}
\def\B{{\cal B}}
\def\C{{\cal C}}
\def\A{{\cal A}}

\def\l{\lambda}
\def\D{\Delta}

\def\h{{1\over 2}}
\def\.{\cdot}
\def\labels#1{\hbox{\hspace{.2in}$_{#1}$} \label{#1}}

\begin{abstract}
An attempt is made
in QCD to explain the growth
of total cross-sections with energy,
without violating the Froissart bound. 
This is achieved by computing the
phase shifts of elastic scatterings
of partons rather than their amplitudes. To 
render that possible a general
formalism of phase-shifts in QCD is developed.
Computed to two-loop order, 
agreements with hadronic
and virtual-photon total cross-sections are quite satisfactory.
Predictions for the slower rate of growth at higher energies
are also presented.
\end{abstract}

The recent observation \cite{zeus,hera} of the
rapid rise of parton density at small $x$
 has generated much theoretical interest
\cite{theory}. This rise is equivalent to
an increase of the total photon-proton cross-section
$\sigma_T(\gamma^*p)$ with its c.m.~energy $\sqrt{s}$. For
photons with virtuality $Q^2$ in the range of the HERA
data, $\sigma_T(\gamma^*p)\sim s^a$
 with the power $a$ substantially larger than 
the power 0.08 \cite{dl}
observed in hadronic and real-photon-hadron 
total  cross-sections, but substantially less than the power
$\sim 0.5$ given by the BFKL Pomeron \cite{bfkl}. 
See Fig.~2. A power growth violates the Froissart bound
$\sigma_T\sim (\ln s)^2$ so eventually all these 
increases have to slow down. 
In terms of parton density, this means
saturation at small $x$ 
when the partons begin to overlap in transverse
dimensions \cite{gribov}. 

A considerable amount of work 
exists in computing parton densities at small $x$ for large
nuclei, and in understanding its saturation
\cite{mueller,mclerran}. We shall approach this problem 
by looking at the 
total cross-section $\sigma_T(\gamma^*p)$,  which at least for 
$Q\gg\Lambda_{QCD}$ is also amenable to  
perturbative QCD calculations. At sufficiently high energies, 
the energy variation of total cross sections
is expected to be universal, so for simplicity
we shall only study the energy dependence of quark-quark
interactions.

Optical theorem relates total 
cross-section to the imaginary part of
forward elastic scattering amplitude. 
If the latter is computed via the exchange of
a BFKL Pomeron \cite{bfkl}, 
then $\sigma_T(s)$ violates the
Froissart bound \cite{bfkl,gribov}. Multi-reggeon exchanges are necessary
to restore unitarity but that calculation is very difficult \cite{reggeon}.
We take a different approach by looking at the 
phase shift, which will almost certainly guarantee
the Froissart bound.
The idea of using phase shift is not new \cite{yang}, 
though it remains a challenge 
to calculate it within the framework of QCD.
We shall discuss how this can be carried out, then  use it
to compute the quark-quark scattering phase-shift to two-loop order,
in the leading-log approximation.
The result compares well with the energy increase of the HERA data
at different $Q^2$ (Fig.~2),
and it also agrees with the hadronic 
data up to laboratory energy of 
$10^9$ GeV (Fig.~3). 

Let $A(s,\Delta)$ be the elastic scattering amplitude
at momentum transfer $\Delta$. At high energies
it is well described by the impact-parameter representation
\be
A(s,\Delta)=is\int d^2b e^{i\vec\Delta\.\vec b}\(1-e^{2i\delta(s,b)}\),
\label{impact}\ee
where $\vec \Delta$ and $\vec b$ are two-dimensional 
vectors in the transverse plane.
At large impact parameters the phase shift $\delta(s,b)$ is small and
it goes approximately like $a\exp(-\mu b)$ if the interaction range
is $\mu^{-1}$. Only for $b$ less than 
the effective radius $R(s)$ does the phase shift contribute
substantially to the amplitude, so $R$ can be
estimated from the condition $\delta(s,R)\sim 1$.
 If $a$ is an increasing function
of $s$, then $R(s)$ and hence $\sigma_T\sim \pi R(s)^2$ 
also increase with $s$. In particular, if $a\sim s^\ell$ 
for $\ell>0$, then $R(s)\sim (\ell/\mu)\ln s$, and $\sigma_T\sim (\ln s)^2$
reaches the Froissart bound. 
This conclusion is 
stable and robust, and is
qualitatively independent of the magnitude of $\ell$ and $\mu$. Unless 
$a$ grows faster than a power of $s$, the
 Froissart bound is almost certainly guaranteed when
computed via the phase shift.

Feynman diagrams tell us how to calculate perturbative amplitudes
but not directly the phase shifts. This is because
the impact-space amplitude
$\A(s,b)=e^{2i\delta(s,b)}-1=
-\sum_{n=1}^\infty[2i\delta(s,b)]^n/ n!$
is given by an infinite sum of all powers of the phase
shift. Even if $\delta(s,b)$ is computed to
the lowest order, the resulting $A(s,\D)$
 already contains terms of
all orders. Thus a proper understanding of phase
shifts cannot be obtained unless an infinite sum of
Feynman diagrams is considered.
More generally, if the phase shift is expanded in powers of the coupling constant
$g^2$, $\delta(s,b)\simeq \sum_mg^{2m}\delta^{(m)}$, then from (\ref{impact})
the $(2n)$th order contribution to $\A(s,b)$ is given by a sum of products
of the phase shifts $\delta^{(m)}$. Individual Feynman diagrams certainly
do not factorize in this manner so even at a given order we have to sum over
many Feynman diagrams. A hint for what to sum is given by 
the formula $\delta(s,b)\sim\int V(s,b,z)dz$ for potential scattering, where $\delta(s,b)$ is given by an accumulation
of interactions (vertices) along the path without
involving any
 energy denominators or propagators. Recall that
in perturbation theory the energy denominator $1/\D E$
comes from the uncertainty relation, and the fact that
the time $\D t\sim
1/\D E$ allowed for each interaction in a given Feynman diagram
is constrained by the times of its two neighbouring interactions. 
To get rid of the energy
denominator we must relax the time constraint
by summing over diagrams with all possible orderings of the interactions. 
When this is carried out with the help of the eikonal formula \cite{eik}
for potential scattering and for QED,
indeed the energy denominators disappear, factorization occurs,
and the phase shift is given by the first Born approximation.

For QCD, the presence of non-commuting colour matrices makes factorization
considerably more difficult to achieve, and 
representations in terms of phase shifts harder to
obtain. In what follows, we shall first
discuss how this is done 
for the simpler case of scattering from an external colour source, before tackling
the more difficult problem of quark-quark scattering
where the leading-log approximation has to be invoked.

The desired factorization for a fast parton of momentum $p$
interacting with external colour sources, 
 in the tree approximation, comes from the decomposition
theorem
 \cite{LL1,Lmp}
\be 
\A_n(\vec b)&=&
\sum_{\{m\}}\C_{m_1}(\vec b)\C_{m_2}(\vec b)\cdots\C_{m_k}(\vec b)
/\prod_{i=1}^{k-1}\sum_{j=i}^km_j,
\label{irred}\ee
with the first sum   taken over all $m_i\ge 1$
so that $\sum_{i=1}^km_i=n$, and over all $k$. For $n>1$,
the {\it irreducible
amplitude} $\C_n(\vec b)$ is identical to the Feynman amplitude
$\A_n(\vec b)$, except the
product of colour matrices $\l_{a_1}\l_{a_2}\cdots\l_{a_{n-1}}\l_{a_n}$
in $\A_n$ is replaced by its nested commutators
$[\l_{a_1},[\l_{a_2},[\cdots,[\l_{a_{n-1}},\l_{a_n}]]]]$ in $\C_n$.
For $n=1$, by definition $\C_1=\A_1$.
Note that $\A_n$ is given by a product of colour matrices so
it contains many colours, but $\C_n$ is given by their nested commutators
so it carries colour only in the adjoint representation, same as the
colour of a single gluon.

With this factorization it is possible to sum
up $\A_n(\vec b)$ to obtain an impact-parameter
representation \cite{Lmp}, from which the formula for the phase shift
can be extracted. The phase shift defined in (\ref{impact}), with
$s$ replaced by $2p$, is given by
\be
2\delta(\vec b)=\sum_{k=1}^\infty {\C_k\over k}+
{1\over 12}[\C_2,\C_1]+{1\over 12}[\C_3,\C_1]+\cdots,
\label{ps}\ee
where the ellipses consists of commutators of $\C_m$ of order 
$g^{10}$ and above.
The $O(g^{10})$ expression is explicitly known \cite{Lmp} and the higher
order expressions can be computed when they are needed.

For quark-quark scattering the presence
of loops and colour matrices of the other quark
makes the problem so much more difficult
that we find it necessary to invoke the leading-log approximation.
It is cruicial to recognize however that the leading-log approximation used here
may not be the same as what is found in the literature.
For example, it is known that the BFKL Pomeron \cite{bfkl}
computed in the leading-log approximation
violates the Froissart
bound; subleading-logs supplied by multiple-reggeon
exchanges are needed to restore unitarity.
This however refers to leading and subleading logs with respect to a
{\it fixed $t$-channel colour} (singlet in this case).
The leading-log approximation to be used in the phase shift calculation keeps
leading-log terms with respect to {\it fixed colour structures}. 
Leading logs in fixed colour structures may give rise to subleading
logs in fixed $t$-channel colours, like those supplied by multiple reggeons,
which is why leading-log for fixed colour structures may give
rise to phase shifts that preserves the Froissart bound.
Indeed, there is a very close relationship 
between colour structures and reggeon structures that will be discussed
further below.

The {\it colour structure} of a diagram is defined by its colour factor 
when quarks carry arbitrary $SU(N_c)$ colours.
Colour factors are 
linearly dependent only if they can be related through $SU(N_c)$
commutation relations alone, with coefficients
 independent of
$N_c$ and the specific quark colours.
The six diagrams in Fig.~1 have linearly independent colour 
structures.

The colour structure of a diagram 
is said to be {\it primitive} if
all the lines in the diagram remain connected after the top and the bottom quark lines
are removed. Using $G$ to denote the colour matrices of a colour 
structure, $G_a$ and $G_d$ in Fig.~1 are primitive and all the others
are not. 

Decomposing gluons attached to one of the quark lines using (\ref{irred}), it can be shown 
in the {\it leading-log approximation} \cite{fl,Lir} that 
(i).~Non-primitive colour factors may be considered
as commutative products of primitive colour factors.
For example, using $G$ to denote the colour 
structure of a diagram, then $G_b=G_a^2$,
$G_c=G_a^3$, $G_e=G_f=G_d^2$; 
(ii).~This colour factorization also leads to an amplitude factorization 
in the impact-parameter space.
If the impact-parameter amplitude for a primitive colour
structure $G_i$ is $d_i$, then the spacetime amplitude for the colour
factor $M(G_i)$, where $M$ is a monomial function of the $G_i$'s,
can be factorized into $M(d_i)$; (iii).~As a result of this factorization,
diagrams of all orders can be summed up to an impact-parameter 
representation with a phase shift given by
$2\delta(s,b)=\sum_i d_iG_i$, where the sum is taken over all 
linearly independent primitive colour factors $G_i$.

To two-loop order in the leading-log approximation, the quark-quark scattering amplitude $A(s,\D)$
is known to be \cite{cw,fhl}
\begin{eqnarray}
&&A(s,\D)/s=\big[-g^2I_1(\D)+(g^4N_c\ln s/4\pi)I_2(\D)\nn
&&-(g^6N_c^2\ln^2s/32\pi^2)\D^2I_2^2(\D)\big]G_a\nn
&&+i\[g^4I_2(\D)/2-g^6N_c\ln sI_3(\D)/4\pi\]G_b\nn
&&+\[g^6I_3(\D)/6\]G_c\nn
&&+i(g^6\ln s/2\pi)\[I_3(\D)-\D^2I_2^2(\D)/2\]G_d,\nn
&&I_n(\D)\equiv\int\prod_{i=1}^n\({d^2k_{i\perp}\over(2\pi)^2
(k_{i\perp}^2+\mu^2)}\)
(2\pi)^2\.\nn
&&\quad\delta^2\(\D-\sum_{i=1}^nk_{i\perp}\)
=\int d^2b\ e^{i\vec\D\.\vec b}\[K_0(\mu b)/2\pi\]^n.
\label{g6}\ee
The infrared cutoff $\mu$ is introduced as a parameter to simulate
hadronic size and confinement \cite{lattice}.
The factorization property mentioned above can be
explicitly verified in eq.~(\ref{g6}). The resulting phase shift to two-loop order
is then given by
\be
2\delta(s,b)&=&2\delta_a(s,b)G_a+2\delta_d(s,b)G_d,\nn
2\delta_a(s,b)&=&-g^2K_0(x)/2\pi+g^4N_c\ln sK_0^2(x)/16\pi^3\nn
&-&g^6N_c^2\ln^2sV(x)/32\pi^2,\nn
2\delta_d(s,b)&=&ig^6\ln s\[K_0^3(x)/16\pi^4-V(x)/4\pi\],
\quad(x\equiv\mu b)\nn
\D^2I_2^2(\D)&\equiv&\int d^2b\ e^{i\vec\D\.\vec b}V(\mu b).\label{ps6}\ee
The colour factors are respectively $G_a=\lambda_a\times\lambda_a$ and $G_d=
\lambda_a\lambda_b\times\lambda_c\lambda_d(if_{ace})(if_{bed})$, where
$\times$ indicates 
tensor product of colour matrices associated with the 
two quark lines. To obtain the elastic amplitude
and hence the total cross section, we must extract
from (\ref{impact}) the contribution of the
 colour-singlet.
The resulting formula for the total cross section 
at $N_c=3$ is
\newpage
\be
\sigma_T(s)={1\over\mu^2}\int d^2x\bigl[1&-&{2\over 3}e^{-D}
\cos\({2\delta_a\over 3}\)\nn
&-&{1\over 3}e^{-2D}
\cos\({4\delta_a\over 3}\)\bigr],\label{final}\ee
where the damping exponent $D=-i\delta_d$ is real and so is
$\delta_a$. Note that these two functions depend on the impact parameter
only through the combination $x=\mu b$, so that all the $\mu$ dependences
are factored out into the $1/\mu^2$ factor in front of the integral.
In this way the uncalculable parameter $\mu$
affects only the overall magnitude of the cross-section, 
but not its energy dependence which is expected to be universal
(as is experimentally the case for hadronic and real photon 
total cross-sections). We may therefore
use this formula for quark-quark
total cross-section on the energy dependence of
experimentally accessible beams and targets.

To compare the prediction of (\ref{final}) with experiment we have to
realize that perturbative QCD contains no energy scale per se, so
an energy scale $\Lambda(Q)$ must be introduced externally to replace
all $\ln s$ factors by $\ln(s/\Lambda^2(Q))$. Recall that 
(\ref{ps6}) is obtained in the leading-log approximation, where
$\ln s$ comes from an integral of the form
$\ln (s/\Lambda^2)=\int_{\Lambda^2/s}d\omega/\omega$, with $\Lambda$
being a combination of the other energy scales present
(masses, momentum transfers, virtualities). For massless quarks and
forward amplitude, on dimensional grounds it is therefore reasonable
to assume $\Lambda(Q)=\Lambda_0+cQ$, with $\Lambda_0$ 
to be of the order
of $\Lambda_{QCD}$, which we shall take it to be 
$0.2$ GeV, and a parameter $c$ which must be determined
phenomenologically. We shall also take the coupling constant to be
$\alpha_s=g^2/4\pi=0.26$, which is its value at $Q^2=0.65$
(GeV$/c)^2$, the largest
$Q^2$ involved in these data. The dependence
on this parameter is very weak so its actual value is really not that
important. In  particular, we will not let it vary with $Q^2$
though there is no problem to include such a variation.
As mentioned before, the parameter $\mu$ simulates confinement and
hadronic-size effects so it is not computable within
the framework of perturbative QCD so it must be fitted for each $Q$.

Total cross sections for deep-inelastic
\cite{zeus} and hadronic \cite{pp} data are shown
in Figs.~2 and 3 respectively. The energy variations predicted by
(\ref{ps6}) and (\ref{final}) are given by the solid curves, with
$c=4$. No attempt has been made to obtain a best fit. 
Shown for comparison in Fig.~2 are also the dotted lines
representing a power variation $s^{0.08}$
and a dashed line giving $s^{0.5}$. In Fig.~3 the dashed line
represents the Donnachie-Landshoff fit \cite{dl,pp} 
$\sigma_T=22s^{0.079}+56.1s^{-0.46}\ mb$.

At $s$ values above a few tens of (GeV)$^2$ for the deep-inelastic
data, and perhaps a bit higher for the hadronic data, the {\it energy
variation} of this theory,
with essentially one free parameter $c$, agrees well with the data.
The $s$-variation of this theory is slower than a power, as it must be
to obey the Froissart bound, but at this energy range it is really not
very different from a power as can be seen in the plots. This theory
with its leading-log approximation is not designed to work at low
energies, but it is amusing to note from Fig.~3 that the dotted line
with the low-energy correction term $s^{-0.46}$
neglected will also come to very similar values at $s\sim 10^2$ (GeV)$^2$.

To summarize, we have suggested that the energy dependence of total 
cross sections should be understood through the phase shifts of the 
elastic amplitudes, because the Froissart bound is almost certainly 
guaranteed in such an approach. The problem of how to calculate
the phase shifts in QCD at high energies is discussed,
exactly for partons scattered from external sources, and in
the leading-log approximation for parton-parton scatterings. 
The latter is also explicitly calculated to two-loop order and
compared with experiments. With an energy scale $\Lambda(Q)=
\Lambda_0+cQ$
this theory compares well with the energy variations of
both the deep inelastic and the
hadronic data.

This research is supported by the Natural Sciences and Engineering 
Research Council of Canada. I wish to thank Fran\c cois Corriveau
for help in obtaining the experimental data and Greg Mahlon for 
instructing me how to plot them.

\newpage

\begin{figure}[h]
\vspace*{5cm}
\includegraphics{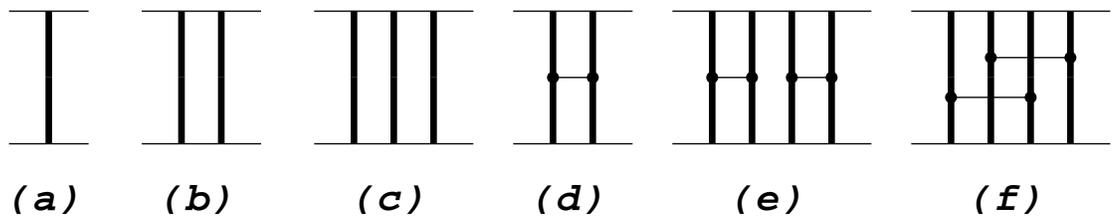}
\vspace*{6cm}
\caption[]{Colour structure diagrams.}
\end{figure}

\newpage

\begin{figure}[h]
\includegraphics{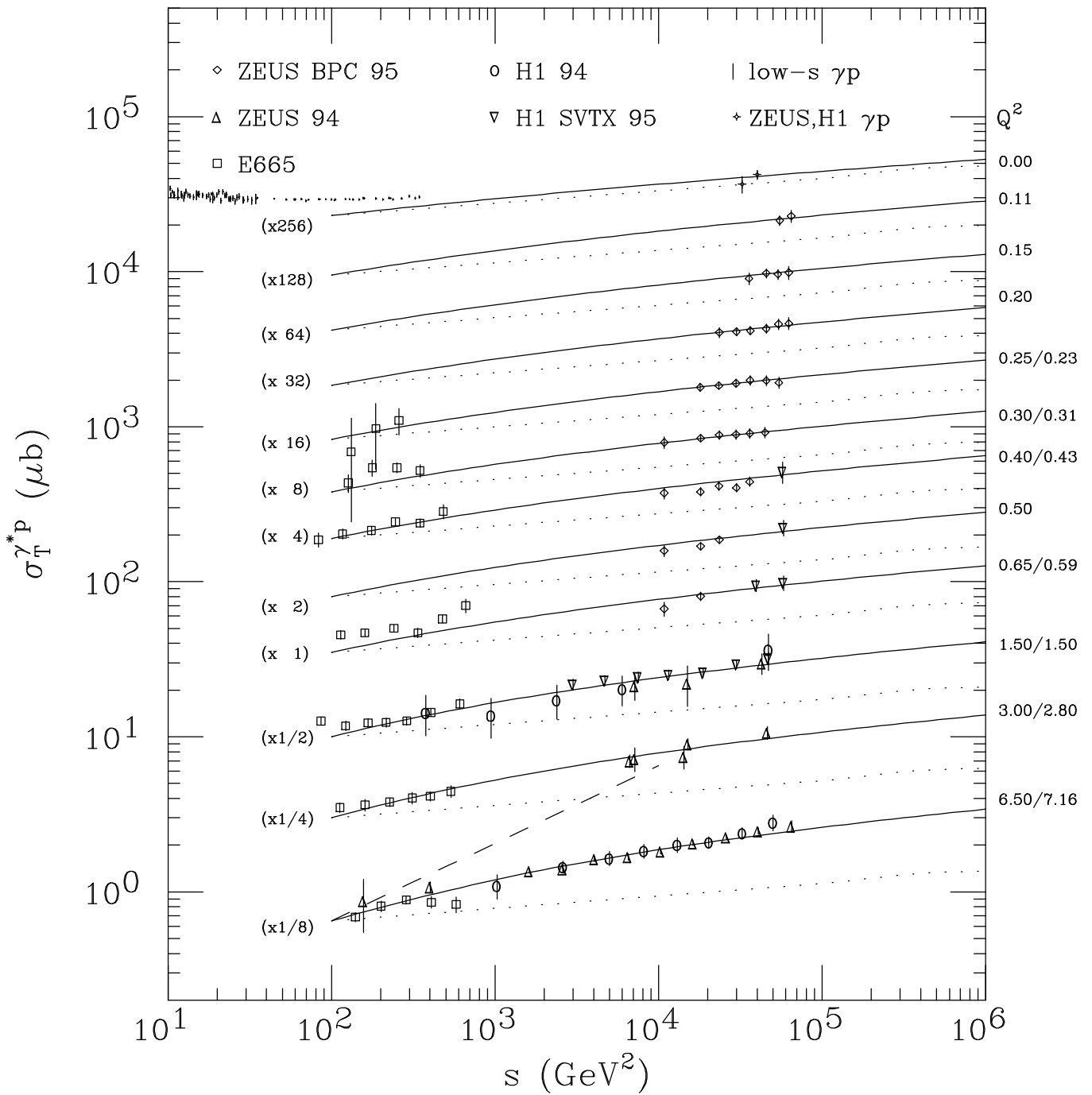}
\vspace*{18.5cm}
\caption[]{
$\gamma^*$-proton total cross-sections as a function of $s$.
Data are taken from Ref.~\protect\cite{zeus} where  references 
to the original experiments can be
found. $Q^2$ of the photon, in (GeV)$^2$, are given on the right.
Where two numbers are listed, the second one refers to the low-energy
data of E665. The solid curve is the prediction of the present theory.
The dotted curve and the dashed curve depict 
respectively energy variations of
$s^{0.08}$ and $s^{0.5}$.}
\end{figure}

\newpage

\begin{figure}[h]
\vspace*{2cm}
\includegraphics{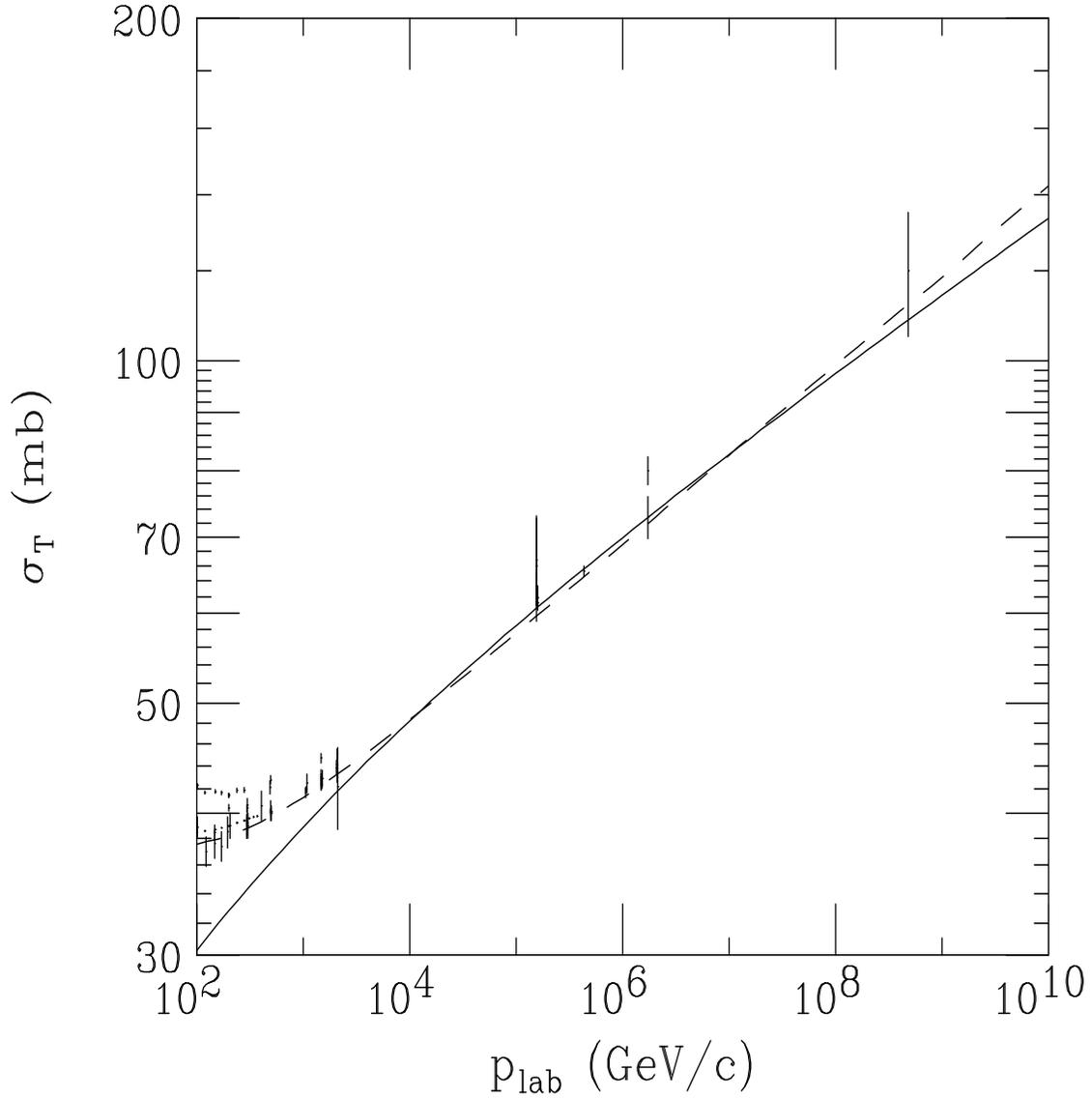}
\vspace*{15cm}
\caption[]{$pp$ and $\bar pp$ total cross sections as a function of 
the laboratory momentum $p_{lab}$. 
Data are taken from Ref.~\protect\cite{pp}, dashed
 line is the Donnachie-Landshoff
fit, and the solid line is the prediction of the present theory.
All parameters are identical to those used in Fig.~2.}
\end{figure}

\end{document}